\documentclass[aps,prl,twocolumn,groupedaddress]{revtex4}

\usepackage{graphicx} 
\usepackage{dcolumn}
\usepackage{bm}

\begin{document}
\title{Consistent explanations of  tunneling and photoemission data in cuprate superconductors: No evidence for magnetic pairing } 
\author{Guo-meng Zhao$^{1,a}$ and A. S. Alexandrov$^{2,3,b}$} 
\affiliation{$^{1}$Department of Physics and Astronomy, 
California State University, Los Angeles, CA 90032, USA~\\
$^{2}$Department of Physics, Loughborough University, Loughborough, LE11 3TU,
U.K.~\\
$^{3}$ Instituto de Fisica ``Gleb Wataghin'', Universidade Estadual de Campinas, UNICAMP 
13083-970, Campinas, S÷ao Paulo, Brazil}

\begin{abstract}

We have analyzed scanning tunneling spectra of two electron-doped cuprates
Pr$_{0.88}$LaCe$_{0.12}$CuO$_{4}$ ($T_{c}$ = 21~K and  24 K) and
compared them with tunneling spectrum of hole-doped La$_{1.84}$Sr$_{0.16}$CuO$_{4}$
and effective electron-boson spectral function 
of hole-doped La$_{1.97}$Sr$_{0.03}$CuO$_{4}$ (extracted from
angle-resolved photoemission spectrum). We have also analyzed
tunneling spectra and angle-resolved photoemission spectra for
hole-doped Bi$_{2}$Sr$_{2}$CaCu$_{2}$O$_{8}$. These results unambiguously rule out magnetic pairing mechanism in both electron- and hole-doped cuprates and support polaronic/bipolaronic superconductivity in hole-doped Bi$_{2}$Sr$_{2}$CaCu$_{2}$O$_{8}$.

\end{abstract}
\maketitle 
Developing the microscopic theory for high-$T_{c}$ superconductivity
is one of the most challenging problems in condensed
matter physics. Twenty-seven years after the discovery of
high-$T_{c}$ superconductivity by Bednorz and M\"uller \cite{Muller},
no consensus on the microscopic pairing mechanism has been reached despite 
tremendous experimental and theoretical efforts. There are essentially two 
opposite views about the pairing mechanism. Many researchers 
believe that antiferromagnetic fluctuations predominantly mediate the electron 
pairing \cite{Review1}. In contrast, other researchers  insist that strong 
electron-phonon coupling is mainly responsible for high-temperature superconductivity in cuprates. 
The polaron-bipolaron theory of superconductivity
\cite{Review2}, which is based on strong electron correlation and
significant electron-phonon interaction (EPI), has gained strong support 
from various experimental results.  In particular, extensive studies of unconventional oxygen-isotope effects in 
hole-doped cuprates have clearly shown strong EPI and the existence of 
polarons/bipolarons in both normal and superconducting states
\cite{ZhaoAF,ZhaoYBCO,ZhaoLSCO,ZhaoNature97,ZhaoJPCM,Zhaoreview1,Zhaoreview2,Zhaoisotope,Keller1,NJP}.
Neutron scattering 
\cite{McQueeney,Sendy,Rez}, angle-resolved photoemission (ARPES) 
\cite{Lanzara01,Ros}, pump-probe \cite{Gad}, and  optical specroscopies \cite{Mih,Mis,YLee,Mis08} have
also demonstrated strong electron-phonon 
coupling. Further, ARPES data \cite{Zhou04} and  tunneling spectra
\cite{Ved,Shim,Gonnelli,Zhao07,Boz08} have consistently
provided direct evidence for strong coupling to multiple-phonon modes 
in hole-doped cuprates.

On the other hand, many researchers still maintain the $d$-wave magnetic pairing 
mechanism, allegedly supported by some highly publicized experimental papers
\cite{Nature,NaturePhysics1,NaturePhysics2}. In one of the papers \cite{NaturePhysics1}, the authors have 
used an unrealistic parameter, which overestimates the magnetic coupling constant by 
two orders of magnitude \cite{ZhaoPS}.   Other two papers \cite{Nature,NaturePhysics2} reported the combined 
neutron and tunneling data for two electron-doped Pr$_{0.88}$LaCe$_{0.12}$CuO$_{4-y}$   
(PLCCO) crystals with different superconducting transition temperatures (21 and 24 K). These data 
seemingly suggest that the energies of the magnetic resonance 
modes are the same as those of the bosonic modes revealed in the second derivative 
(d$^{2}I$/d$V^{2}$) of electron tunneling current ($I$) with respect to bias voltage
($V$). 
They thus conclude that the magnetic resonance modes rather than phonons mediate electron 
pairing in electron-doped cuprates \cite{Nature,NaturePhysics2}. However, 
one of us (GMZ) \cite{Zhao09} has already pointed out a basic mistake in the 
data analyses of Ref.~\cite{Nature} and shown that the combined neutron and 
tunneling data actually disprove this magnetic pairing mechanism. Despite 
citing Ref.~\cite{Zhao09}, the same basic mistake has been repeated in Ref.~\cite{NaturePhysics2}. Based on the 
repeated incorrect analyses, these authors concluded that their data support $d$-wave magnetic 
pairing mechanism.

Here we re-analyze scanning tunneling spectra of two electron-doped cuprates
Pr$_{0.88}$LaCe$_{0.12}$CuO$_{4}$ ($T_{c}$ = 21~K and  24 K)  and
compare them with tunneling spectrum of hole-doped La$_{1.84}$Sr$_{0.16}$CuO$_{4}$
and effective electron-boson spectral function $\alpha{^2}(\omega)$$F(\omega)$
of hole-doped La$_{1.97}$Sr$_{0.03}$CuO$_{4}$. Our data analysis along with 
other independent tunneling and ARPES data of hole-doped Bi$_{2}$Sr$_{2}$CaCu$_{2}$O$_{8}$ (BSCCO) consistently rules out  magnetic 
pairing mechanism in both electron- and hole-doped cuprates and supports polaronic/bipolaronic superconductivity in hole-doped Bi$_{2}$Sr$_{2}$CaCu$_{2}$O$_{8}$. 

\begin{figure}[htb]
    \includegraphics[height=11cm]{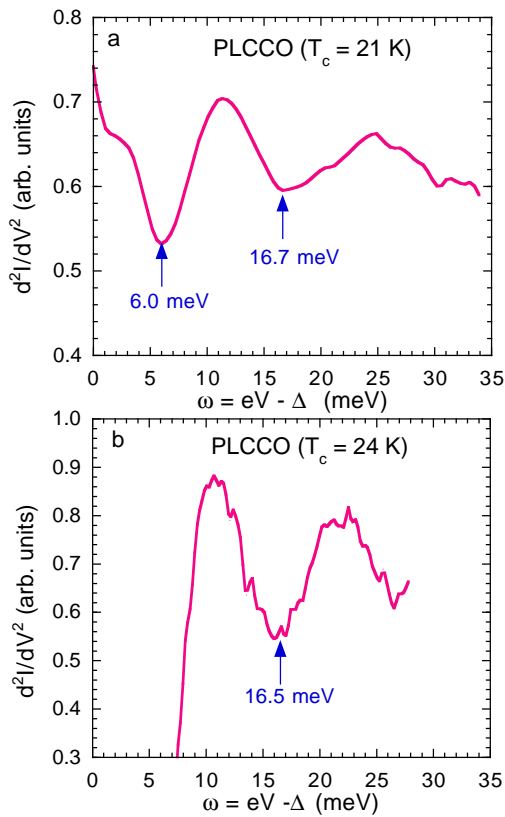}
 \caption[~]{a) d$^{2}I$/d$V^{2}$ spectrum for an electron-doped PLCCO sample with $T_{c}$ = 21 K, which  is reproduced from Ref.~\cite{NaturePhysics2}.  b) d$^{2}I$/d$V^{2}$ spectrum for an electron-doped PLCCO sample with $T_{c}$  = 24 K, which is reproduced from Ref.~\cite{Nature}. The energy positions of the d$^{2}I$/d$V^{2}$ spectra  are measured from the superconducting gaps. }
\end{figure}

It has been well established \cite{Shim,Zhao07,Boz08,MR,Carb}) that the energies of the dip positions 
(rather than the peak positions) in d$^{2}I$/d$V^{2}$ correspond to the energies of 
the modes strongly coupled to electrons. In Fig.~1, we adopt this well-established 
protocol to identify the mode energies from the d$^{2}I$/d$V^{2}$ spectra of two 
electron-doped samples. For the electron-doped sample with $T_{c}$ = 21 K, the mode 
energies identified from the d$^{2}I$/d$V^{2}$ spectrum below 35 meV are 6.0 meV 
and 16.7 meV (indicated by the arrows). For the electron-doped sample with $T_{c}$ = 24 K, 
the mode energy identified from the d$^{2}I$/d$V^{2}$ spectrum between 7 and 28 meV 
is 16.5 meV.
It is apparent that the mode energies revealed in the d$^{2}I$/d$V^{2}$ spectra of 
the two electron-doped cuprates are nearly the same (16.7 and 16.5 meV) and significantly 
different from the magnetic resonance energies (9.0 or 10.5 meV) revealed by inelastic 
neutron scattering experiments \cite{Nature,NaturePhysics2}. The mode energy of about 16.5 meV, which is independent of 
doping and $T_{c}$, agrees with the energies 
of the two transverse optical (TO) phonon modes (15.6 meV for the $E_{u}$ mode and 17.0 meV 
for the $A_{2u}$) \cite{Crawford}. This implies that the phonon modes rather than the magnetic-resonance 
modes mediate electron pairing in electron-doped cuprates. This is also consistent with both ARPES data \cite{Mee1} of hole-doped
 Bi$_{2}$Sr$_{2}$CuO$_{6}$ and theoretical studies \cite{Mee2,Bauer}, which show that 
 the optical phonons are strongly coupled to electrons due to the unscreened
 long-range interaction along the $c$-axis.

\begin{figure}[htb]
    \includegraphics[height=11cm]{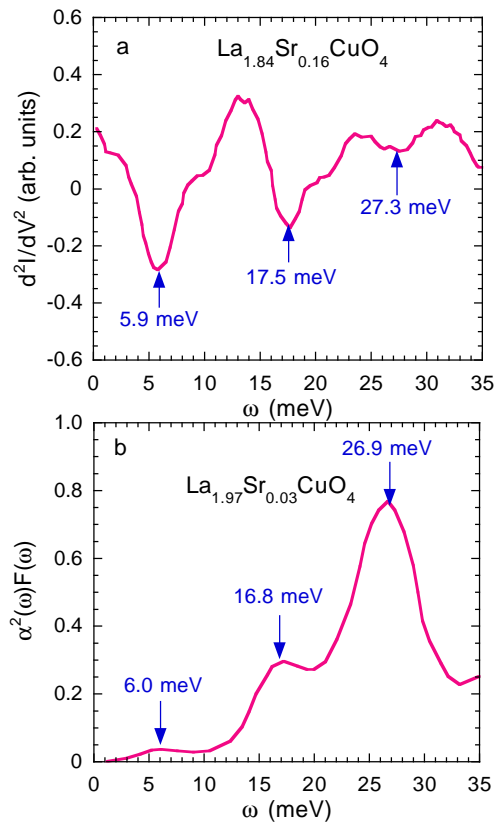}
 \caption[~]{a) d$^{2}I$/d$V^{2}$ spectrum at 7.2 K for hole-doped
 La$_{1.84}$Sr$_{0.16}$CuO$_{4}$, which is 
 reproduced from Fig.~4 of Ref.~\cite{Boz08}. The energy position of the d$^{2}I$/d$V^{2}$ 
 spectrum  is measured from the superconducting gap.  b) Effective electron-boson spectral function $\alpha (\omega)F(\omega)$
 for the hole-doped La$_{1.97}$Sr$_{0.03}$CuO$_{4}$, which is
 extracted from high-resolution ARPES data. The curve is reproduced from Ref.~\cite{Zhou04}. }
\end{figure}

It is interesting that in the structurally similar single-layer hole-doped 
La$_{1.84}$Sr$_{0.16}$CuO$_{4}$ (LSCO) sample, there is a similar mode with energy of 
17.5~meV (see Fig.~2a). The same mode with energy of 16.8 meV is independently revealed 
in the effective electron-boson spectral function $\alpha (\omega)F(\omega)$ of hole-doped 
La$_{1.97}$Sr$_{0.03}$CuO$_{4}$ sample (see Fig.~2b). It is striking that the mode energies
(5.9 meV, 17.5 meV, and 27.3 meV)
inferred from the dip positions of the tunneling spectrum match the peak positions (6.0 meV, 16.8 meV, and 26.9 meV) in the effective 
electron-boson spectral function independently determined  from ARPES 
data \cite{Zhou04}. This further justifies the correctness of our mode assignment.  

Since the magnetic-resonance 
mode energy was found \cite{JZhao} to be proportional to $T_{c}$, insensitivity of 
the mode energies to $T_{c}$ further argues against the magnetic origin of the modes.  
The similar mode energies across different doping levels (Fig.~2) and across different 
single-layer structures (PLCCO and LSCO) are consistent with the phonon mode assignment 
and contradict the magnetic mode assignment. 

\begin{figure}[htb]
    \includegraphics[height=11cm]{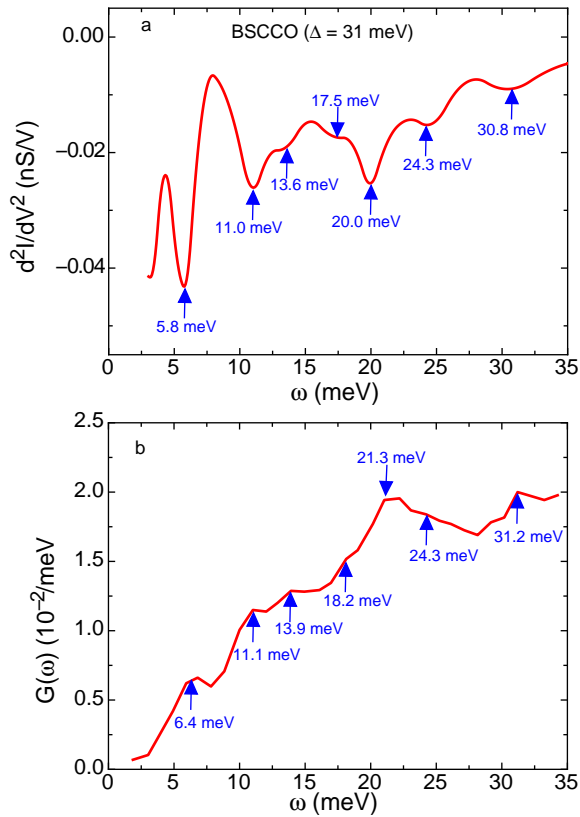}
 \caption[~]{a) d$^{2}I$/d$V^{2}$ spectrum of double-layered hole-doped  Bi$_{2}$Sr$_{2}$CaCu$_{2}$O$_{8}$. The d$^{2}I$/d$V^{2}$ spectrum is obtained by numerically taking the derivative of the d$I$/d$V$ spectrum (positive bias) of Ref.~\cite{Pan} after the spectrum is smoothened by cubic spline.   The energy position of the d$^{2}I$/d$V^{2}$ 
 spectrum  is measured from the superconducting gap ($\Delta$ = 31.0 meV). b) Phonon density of states for slightly overdoped  Bi$_{2}$Sr$_{2}$CaCu$_{2}$O$_{8}$. The curve is reproduced from Ref.~\cite{Renker}. }
\end{figure}

 For double-layered hole-doped  Bi$_{2}$Sr$_{2}$CaCu$_{2}$O$_{8}$, strong coupling features below 35 meV are also seen in the d$^{2}I$/d$V^{2}$ spectrum (see Fig.~3a).  It is striking that the energies of all the dip features in the d$^{2}I$/d$V^{2}$ spectrum precisely match the energies of the peak features in the phonon density of states obtained from high-resolution inelastic neutron scattering \cite{Renker}. This suggests that these bosonic modes strongly coupled to electrons should be the phonon modes. Therefore, strong coupling to multiple phonon modes  is universally seen in single-layered electron- and hole-doped cuprates and in double-layered  hole-doped cuprates.

It is worth noting that the bosonic mode at about 6.0 meV seen in the overdoped $n$-type cuprate (Fig.~1a) 
and 
in the $p$-type La$_{1.84}$Sr$_{0.16}$CuO$_{4}$ (Fig.~2a) and  
La$_{1.97}$Sr$_{0.03}$CuO$_{4}$ (Fig.~2b)
is also seen in the tunneling spectrum of Bi$_{2}$Sr$_{2}$CaCu$_{2}$O$_{8}$ 
(Fig.~3a) and in the tunneling spectrum of YBa$_{2}$Cu$_{3}$O$_{7}$ (Ref.~\cite{Zhao07}), as well as in the ARPES data of Bi$_{2}$Sr$_{2}$CaCu$_{2}$O$_{8}$ 
(Ref.~\cite{Zhao07symmetry,Shen10,Dau10}).  This indicates that this mode is universal for the cuprate systems and
cannot originate from antiferromagnetic fluctuations. 

\begin{figure}[htb]
    \includegraphics[height=11cm]{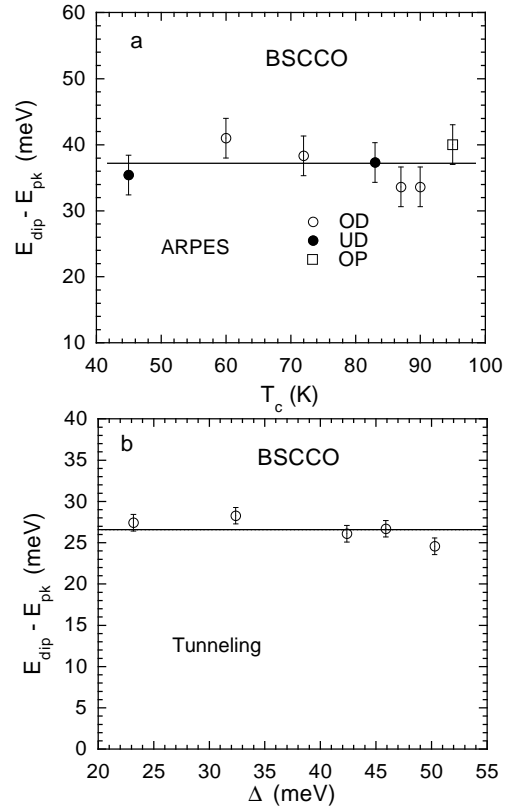}
 \caption[~]{a) Dependence of $E_{dip}-E_{pk}$ on $T_{c}$, which is reproduced from 
 Ref.~\cite{Zhaoreview2}. The $E_{dip}-E_{pk}$ values were determined from ARPES data.  b) Dependence 
 of $E_{dip}-E_{pk}$ on the superconducting gap $\Delta$.  The $E_{dip}-E_{pk}$ values are obtained 
 from Supplementary Figure 2 of  Ref.~\cite{Lee}.  The different $E_{dip}-E_{pk}$ values extracted from ARPES and tunneling data may arise from different energy resolutions that affect peak width.}
\end{figure}

The phenomenological spin-fermion model of Abanov and
Chubukov \cite{Abanov} showed that  strong coupling 
to the magnetic resonance mode yields a peak-dip-hump (PDH) 
structure in ARPES spectra along the antinodal direction \cite{Abanov}.  This theory is based on the one-loop correction to the
$t-t'-J$ mean-field theory. This PDH structure 
was also predicted to be present in d$I$/d$V$ tunneling spectra based on the conventional strong coupling theory \cite{Zasa}.  Within both approaches, the energy separation 
between the dip and peak features is exactly equal to the energy $E_{r}$ of the magnetic resonance 
mode \cite{Abanov,Zasa}, that is, $E_{dip}-E_{pk}$ = $E_{r}$. Since $E_{r}$ was found \cite{JZhao}
to be proportional to $T_{c}$,  $E_{dip}-E_{pk}$ should also be proportional to $T_{c}$. This is in sharp 
contrast to the experimental results shown in Fig.~4. The $E_{dip}-E_{pk}$ 
values obtained from ARPES (Fig.~4a) and 
tunneling spectra (Fig.~4b) are nearly independent of doping or $T_{c}$, in contradiction with the 
the $d$-wave magnetic pairing mechanisms based on the spin-fermion model \cite{Abanov} and on the conventional approach \cite{Zasa}.  Therefore, the results shown in Fig.~4 rule out the magnetic pairing mechanism based on the phenomenological spin-fermion model and on the conventional approach.

More quantitative approach to the $t-t'-J$  model is the  slave-boson
mean-field theory developed by Brinckmann and
Lee \cite{Brink}. The pairing interaction is the antiferromagnetic exchange energy $J$. This mean-field theory also predicts the PDH structure in ARPES spectra along the antinodal direction \cite{Li}. The predicted $E_{dip}-E_{pk}$ is nearly independent of doping and in the range of 0.32-0.46$J$ depending on the energy resolution \cite{Li}. With $J$ = 130 meV,  one yields $E_{dip}-E_{pk}$ = 41.6-59.8 mV, which is in reasonable agreement with the data shown in Fig.~4a. Thus, the result shown in Fig.~4a alone cannot rule out the magnetic pairing mechanism based on the $t-J$ model. 

\begin{figure}[htb]
    \includegraphics[height=5.5cm]{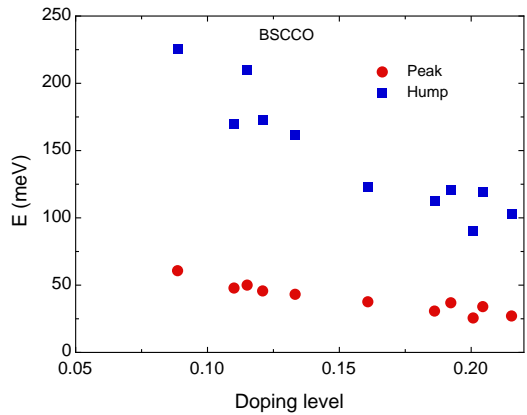}
 \caption[~]{Doping dependence of the energy position of the peak and hump features of the angle-resolved photoemission spectra along the antinodal direction.  The data were reproduced from Ref.~\cite{Camp}. }
\end{figure}

On the other hand, the $t-t'-J$ model predicts a doping independent  $E_{r}$ = 0.54$J$ = 70.2 meV in the overdoped range \cite{Li}. This prediction is in contradiction with neutron data \cite{He}, which show that  $E_{r}$ is equal to 5.4$k_{B}T_{c}$ in the overdoped range.  For overdoped Y$_{0.85}$Ca$_{0.15}$Ba$_{2}$Cu$_{3}$O$_{7}$ with $T_{c}$ = 75 K, $E_{r}$ was found to 34 meV (Ref.~\cite{Pai}), which is only half the predicted value of 70 meV. Therefore, the neutron data rule out the magnetic pairing mechanism based on the $t-t'-J$ model.

The second important prediction of the $t-t'-J$  model is the existence of the pronounced hump feature at energy close to $J$ (about 130 meV) at all the doping levels \cite{Li}. This is in contradiction with the ARPES data shown in Fig.~5. The energy position of the hump feature increases rapidly with the decrease of doping and reaches a value of about 230 meV at a doping level of 0.089. Therefore, the ARPES data further rule out the magnetic pairing mechanism based on the $t-t'-J$ model.

\begin{figure}[htb]
    \includegraphics[height=11cm]{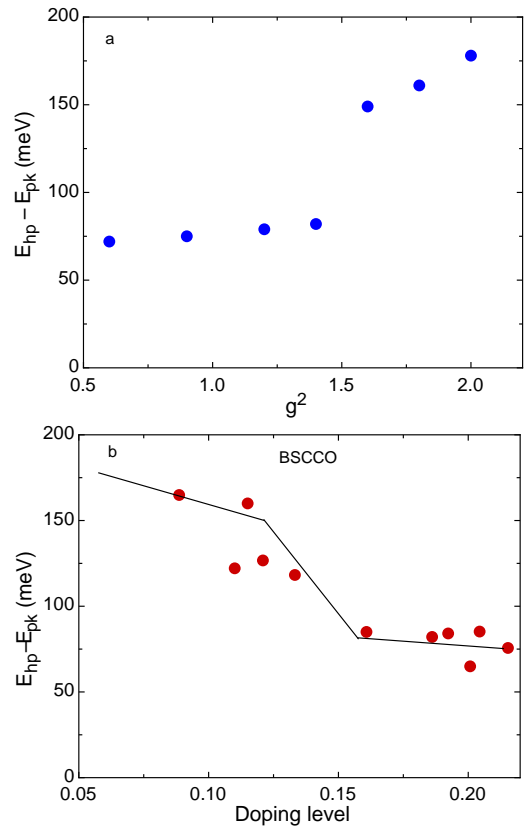}
 \caption[~]{a) Dependence of the peak-hump energy separations on the electron-phonon coupling strength $g^{2}$ in break-junction tunneling spectra, which is predicted from a lattice polaronic model where 72-meV phonon modes are involved in polaron formation. The calculated tunneling spectra are shown in Fig.~1 of the Supplemental Material \cite{SM}.  b) The  peak-hump energy separations for BSCCO, which are calculated from the data in Fig.~5.  }
\end{figure}

Alternatively, the peak-dip-hump features consistently observed in scanning tunneling, break-junction tunneling, and angle-resolved photoemission spectra can be quantitatively explained by the polaron-bipolaron theory of superconductivity, where the hump is the incoherent broad feature that reflects the local hopping of electrons in various frozen
lattice configurations \cite{AlexEPL}. This theory \cite{AlexEPL} predicts that $E_{hp}-E_{pk}$ in break-junction (superconductor-insulator-superconductor) tunneling spectra is close to the energy ($\Omega$) of the phonon modes in polaronic cloud  when the coupling strength $g^{2}$ is not too large. Fig.~6a shows the theoretical prediction of 
$E_{hp}-E_{pk}$ with varying $g^{2}$ and a fixed $\Omega$ = 72 meV. The numerically calculated break-junction tunneling spectra are demonstrated in Fig.~1 of the Supplemental Material \cite{SM}. It is apparent that $E_{hp}-E_{pk}$ is close to $\Omega$ when $g^{2}$ is below 1.5 and close to 2$\Omega$ when $g^{2}$ is above 1.5.  This theoretical prediction is in quantitative agreement with the data shown in Fig.~6b. The data are consistent with $g^{2}$ $<$~1.5 for optimally and overdoped BSCCO and with 1.5~$\leq$~$g^{2}$~$\leq$~2.0 for underdoped BSCCO.

The 72~meV phonon mode should be mainly associated with oxygen vibration so that the whole broad hump feature should shift down by about 72(1-$\sqrt{18/16}$) meV = 4.3 meV upon replacing $^{16}$O by $^{18}$O. Indeed, the energy of the peak feature at 52 meV in d$^{2}I$/d$V^{2}$ spectra was found to shift shown by 3.7$\pm$0.8 meV upon replacing $^{16}$O by $^{18}$O (Ref.~\cite{Lee}). Since the peak position in d$^{2}I$/d$V^{2}$  simply corresponds to the steepest point of the hump feature in d$I$/d$V$, the oxygen isotope shift of the steepest point at 52 meV implies that the hump feature is also shifted down by about 3.7$\pm$0.8 meV upon replacing $^{16}$O by $^{18}$O. Therefore, the observed oxygen-isotope shift of the steepest point of the hump feature in d$I$/d$V$ provides further proof that PDH arises from the lattice polaronic effect.

In summary, our analyses of tunneling and ARPES spectra  in both electron and hole doped cuprates have unambiguously ruled out  magnetic pairing mechanism and support polaronic/bipolaronic superconductivity in hole-doped Bi$_{2}$Sr$_{2}$CaCu$_{2}$O$_{8}$.  
 For detailed discussions about the polaron-bipolaron theory and the intrinsic pairing symmetry in cuprate superconductors, see the Supplemental Material \cite{SM}.

~\\
$^{a}$~gzhao2@calstatela.edu~\\
$^{b}$~The author was sadly deceased right after the manuscript was intially submitted to Physical Review Letters on August 14, 2012.

\bibliographystyle{prsty}

\end{document}